\documentclass[aps,twocolumn]{revtex4}
\usepackage{amsmath}
\usepackage{graphicx}
\begin{document}

\title{Persistence in Advection of Passive Scalar}
\author{
D.Chakraborty \footnote {e-mail:tpdc2@mahendra.iacs.res.in }}
\affiliation{Department of Theoretical Physics,\\
Indian Association for the Cultivation of Science,\\
Jadavpur, Kolkata 700 032, India.}

\begin{abstract}
We consider the persistence phenomenon in advectecd passive scalar equation in 1-dimension. The velocity field is random with the $\langle v(k,\omega)v(-k,-\omega) \rangle \sim |k|^{-(2+\alpha)}$.
In presence of the non-linearity the complete Green's function becomes $G^{-1}=-i\omega+Dk^2+\Sigma$. We determine $\Sigma$ self-consistently from the correlation function which gives $\Sigma \sim k^{\beta}$, with $\beta=(1-\alpha)/2$. The effect of the non-linear term in the equation in the $\mathcal{O}(\epsilon^2)$ is to replace the diffusion term due to molecular viscosity by an effective term of the form $\Sigma_0 k^{\beta}$. The stationary correlator for this system is $[\mathrm{Sech}(T/2)]^{1/\beta}$. Using the self-consistent theory we have determined the relation between $\beta$ and $\alpha$. Finally, IIA is used to determine the persistent exponent.
\end{abstract}
\maketitle

The problem of persistence or survival \cite{1} has attracted a lot of interest in the last decade. The persistence probability has been obtained both analytically and numerically for a large class of stochastic process, Markovian as well as Non-Markovian. The random walk problem, diffusion problem, surface growth, Ising model
with Glauber dynamics are only few to name \cite{1}-\cite{24}. The persistence probability or the zero crossing probability is simply the probability that the local field $\phi(x,t)$ has not changed sign up to time $t$. For single particle systems such as the random walker, which are also Markovian in nature, the persistence probability is easy to calculate since the stationary correlator of such a process decays exponentially. For many body systems where the field $\phi$ has a space dependence the calculation of the zero crossing probability becomes complicated. The problem is now two fold- first we have to write down an effective equation for a single site process by solving the underlying dynamics of the many-particle system and then from this effective single site equation we have to find out the persistence probability. Even though the first part is achieved, the second part of obtaining the persistence probability is notoriously tough since the resulting single site process becomes non-Markovian.

The simplest of such a process which one can think of is the diffusion equation $\partial_t \phi=D\partial^2_x \phi$. The fact that this is effective single site equation can be seen from the solution $\phi(x,t)=\int \mathrm{d}x' G(x-x',t)\phi(x',0)$, where $G(x,t)$ is the Green's function for diffusion equation. The problem of persistence in a diffusion equation has already been addressed by Majumdar et.al \cite{2}. They considered the diffusion equation with random initial condition $\phi(x,0)$ taken from a Gaussian distribution. The two time corrleation function $C(t_1,t_2)$of the normalized variable $X=\phi(x,t)/\sqrt{\langle \phi^2(x,t)\rangle}$ takes the form
\begin{equation}
\label{1}
 C(t_1,t_2)\equiv \langle X(t_1)X(t_2) \rangle =[4t_1 t_2/(t_1+t_2)^2]^{D/4},
\end{equation}
where $D$ is the dimension of space.
Now if we make the transformation $T=\ln t$, the correlator $C(T_1,T_2)$ becomes $f(|T_1-T_2|)$, with $f(T)=[sech(T/2)]^{D/2}$, which is clearly stationary. The stationary correlator for the effective single site process is not exponentially decaying and therefore the calculation of the persistence exponent becomes difficult. The fact that the correlator is not exponentially decaying indicates that the effective single site process is non-Markovian because of the interaction with nearest neghibour sites. Given this stationary non-Markovian correlator it then remains to determine the persistence probability. Two methods have been developed to address this problem, the Independent Interval Approximation (IIA) \cite{2} and the ``series expansion'' \cite{9} approach. In this present article we will use IIA to evaluate the exponents.

Knowing the information about the persistence exponents for a diffusive process, it is natural to ask what would be zero crossing probability when the diffusive process is augmented by an advection term. For a simple diffusive process, if $L$ be the relevant length scale (say the size of the container), then the time to diffuse to a distance $L$ is simply $\tau_1=L^2/2D$. If, however, the particles are advected then the the time for them to diffuse through a distance $L$ is $\tau_2=L/v$, where $v$ is the advection velocity. The ratio of the two time scales is 
\begin{equation}
\label{2}
 \tau_1/\tau_2=2S/Re,
\end{equation}
where $S$ is the Schmidt number and $Re$ is the Reynolds number. The Schmidt number is of the order of unity and therefore, it follows that the mixing time due to advection is smaller than the pure diffusive process. We, therefore expect that the exponents will be greater than those for the pure diffusive process.

The advected passive scalar equation reads
\begin{equation}
\label{3}
 \frac{\partial \phi}{\partial t} + \vec{v}(\vec{x},t)\cdot \nabla \phi = D \nabla^2 \phi
\end{equation}
together with 
\begin{equation}
\label{4}
 \nabla \cdot \vec{v}=0.
\end{equation}

In 1-dimension however, the constraint imposed by Eq.(\ref{4}) is relaxed. Instead, in 1-dimension, we will consider a random velocity field drawn from a given distribution. In particular, the velocity-velocity correlation is given by
\begin{equation}
\label{5}
 \langle v(x,t) v(x',t')\rangle = 2D g(|x-x'|) \delta(t-t').
\end{equation}

The Fourier transform of Eq.(\ref{3}) in one dimension is 
\begin{equation}
\label{6}
 \frac{\partial \tilde{\phi}}{\partial t} +i\int \frac{\mathrm{d} p}{2\pi} p \tilde{v}(k-p,t) 
\tilde{\phi}(p,t)=Dk^2\tilde{\phi}(k,t)
\end{equation}
and the velocity-velocity correlation transforms to
\begin{equation}
\label{7}
 \langle \tilde{v}(k,t)\tilde{v}(k',t)\rangle=2D\tilde{g}(|k|) \delta(k+k')\delta(t-t'),
\end{equation}
where $\tilde{\phi}(k,t)$, $\tilde{v}(k,t)$ and $\tilde{g}(k)$ are the Fourier transform of $\phi(x,t)$, $v(x,t)$ and $g(x)$ respectively. We choose $\tilde{g}(|k|)$ as a power law decaying function, that is,
\begin{equation}
\label{8}
 \tilde{g}(|k|)=\frac{1}{|k|^{(2+\alpha)}}.
\end{equation}
with $0<\alpha<1$.
In absence of the non-linear term the Green's function for Eq.(\ref{6}) is
\begin{equation}
\label{9}
G_0^{-1}=-i\omega+Dk^2.
\end{equation}
The effect of the nonlinearity is to replace the zeroth order Greens function by its most general form
\begin{equation}
\label{10}
G^{-1}=-i\omega+Dk^2+\Sigma=G_0^{-1}+\Sigma.
\end{equation}
Hence,
\begin{equation}
\label{11}
G=\frac{G_0}{1+G_0\Sigma}=G_0(1-\Sigma G_0 +\Sigma^2 G_0^2+......)
\end{equation}
The correlation function $\langle \tilde{\phi}(k,\omega)\tilde{\phi}(-k,-\omega)\rangle$ can be written as
\begin{equation}
\label{12}
\langle \tilde{\phi}(k,\omega)\tilde{\phi}(-k,-\omega)\rangle \sim GG=G_0G_0-G_0\Sigma G_0 +G_0\Sigma^2 G_0^2+.....
\end{equation}
We will determine the self-energy $\Sigma$ self consistently from the correlation function.

Taking a Fourier transformation of Eq.(\ref{6}) in time domain we get
\begin{eqnarray}
\label{13}
\nonumber
-i\omega \tilde{\phi}(k,\omega)+i\int \frac{\mathrm{d}p}{2\pi} \frac{\mathrm{d}\omega'}{2\pi} p \tilde{v}(k-p,\omega-\omega')\tilde{\phi}(p,\omega')\\
=-Dk^2\tilde{\phi}(k,\omega)
\end{eqnarray}

while the velocity corrleation function becomes
\begin{equation}
\label{14}
\langle \tilde{v}(k,\omega)\tilde{v}(k',\omega')\rangle=2D\tilde{g}(|k|) \delta(k+k')\delta(\omega+\omega').
\end{equation}
We now make a perturbative expansion in $\phi$ and write 
\begin{equation}
\label{15}
\tilde{\phi}=\tilde{\phi}_0+\epsilon \tilde{\phi}_1+\epsilon^2 \tilde{\phi}_2+\ldots
\end{equation}
Substituting this in Eq.(\ref{13}), the zeroth order solution is
\begin{eqnarray}
\label{16}
\nonumber
\tilde{\phi}_0(k,\omega)[-i\omega+Dk^2]=\tilde{\phi}_0(k,0)\\
\tilde{\phi}_0(k,\omega)=\frac{\tilde{\phi}_0(k,0)}{[-i\omega+Dk^2]}=G_0(k,\omega)\tilde{\phi}_0(k,0)
\end{eqnarray}
In the first order the solution for $\phi_1(k,\omega)$ is
\begin{equation}
\label{17}
\phi_1(k,\omega)=G_0(k,\omega)[-i\int \frac{\mathrm{d}p}{2\pi}\frac{\mathrm{d}\omega'}{2\pi} \quad p v(k-p,\omega-\omega')\phi_0(p,\omega')]
\end{equation}
while the solution for $\tilde{\phi}_2$ becomes
\begin{eqnarray}
\label{18}
\nonumber
\tilde{\phi}_2(k,\omega)=G_0(k,\omega)[-i\int \frac{\mathrm{d}p}{2\pi}\frac{\mathrm{d}\omega'}{2\pi} p v(k-p,\omega-\omega') \tilde{\phi}_1(p,\omega')]\\
\nonumber
=G_0(k,\omega)\biggr[-i\int \frac{\mathrm{d}p}{2\pi}\frac{\mathrm{d}\omega'}{2\pi} p v(k-p,\omega-\omega')G_0(p,\omega')\\
\nonumber
\biggr\{-i\int \frac{\mathrm{d}q}{2\pi}\frac{\mathrm{d}\omega''}{2\pi} qv(p-q,\omega'-\omega'')\tilde{\phi}_0(q,\omega'')\biggr\}\biggr]\\
\end{eqnarray}
To evaluate $\Sigma$ self consistently we need to calculate the corrleation function $\langle \tilde{\phi}_1(k,\omega)\tilde{\phi}_1(-k,-\omega)\rangle$
$\langle \tilde{\phi}_2(k,\omega)\tilde{\phi}_0(-k,-\omega)\rangle$.
We assume that the non-linear contribution to the total Green's function $G$ will dominate over the $Dk^2$ term \cite{25}. Hence, we rewrite Eq.(\ref{10}) as 
\begin{equation}
\label{19}
G^{-1}=-i\omega+\Sigma,
\end{equation}
which shows that $\omega$ and $\Sigma$ have the same dimension.

The correlation $\langle \tilde{\phi}_2(k,\omega)\tilde{\phi}_0(-k,-\omega)\rangle$ is then
\begin{eqnarray}
\label{20}
\nonumber
\langle \tilde{\phi}_2(k,\omega)\tilde{\phi}_0(-k,-\omega)\rangle=-G_0(k,\omega)\langle\biggr[ \int \frac{\mathrm{d}p}{2\pi}\frac{\mathrm{d}\omega'}{2\pi}\\
\nonumber
p v(k-p,\omega-\omega')G_0(p,\omega')\\
\nonumber
\biggr\{-i\int \frac{\mathrm{d}q}{2\pi}\frac{\mathrm{d}\omega''}{2\pi} qv(p-q,\omega'-\omega'')\tilde{\phi}_0(q,\omega'')\biggr\}\tilde{\phi}_0(-k,-\omega)\biggr]\rangle\\
\end{eqnarray}
A little algebra simplifies the above expression to 
\begin{eqnarray}
\label{21}
\nonumber
\langle \tilde{\phi}_2(k,\omega)\tilde{\phi}_0(-k,-\omega)\rangle=-G_0(k,\omega)\biggr[ \int
\frac{\mathrm{d}p}{2\pi}\frac{\mathrm{d}\omega'}{2\pi}\frac{\mathrm{d}\omega''}{2\pi} \\
\nonumber
kp\langle v(k-p,\omega-\omega') 
v(p-k,\omega'-\omega'')\rangle G_0(k,\omega'')G_0(-k,-\omega)\biggr ].\\
\end{eqnarray}
The velocity-velocity correlation gives a $\delta(\omega-\omega'')$ which, after the $\omega''$ integral becomes
\begin{eqnarray}
\label{22}
\nonumber
\langle \tilde{\phi}_2(k,\omega)\tilde{\phi}_0(-k,-\omega)\rangle=-G_0(k,\omega)\biggr[\int \frac{\mathrm{d}p}{2\pi}\frac{\mathrm{d}\omega'}{2\pi}\frac{kp}{|k-p|^{(2+\alpha)}}\\
\nonumber
G_0(p,\omega')\biggr]G_0(k,\omega)G_0(-k,-\omega)\\
\end{eqnarray}

We now turn our attention to $\langle \tilde{\phi}_1(k,\omega)\tilde{\phi}_1(-k,-\omega)\rangle$ which is given by
\begin{eqnarray}
\label{23}
\nonumber
\langle \tilde{\phi}_1(k,\omega)\tilde{\phi}_1(-k,-\omega)\rangle=-G_0(k,\omega) \biggr[ \int \frac{\mathrm{d}p}{2\pi}\frac{\mathrm{d}\omega'}{2\pi}\frac{\mathrm{d}q}{2\pi}\frac{\mathrm{d}\omega''}{2\pi}\\
\nonumber
pq \langle v(k-p,\omega-\omega')v(-k-q,-\omega-\omega')\rangle \\
\nonumber
\langle \tilde{\phi}_0(p,\omega')\tilde{\phi}_0(q,\omega'')\rangle\biggr] G_0(-k,-\omega)\\
\end{eqnarray}
The velocity-velocity correlation introduces a $\delta(p+q)\delta(\omega'+\omega'')$ while the average $\langle \tilde{\phi}_0(p,\omega')\tilde{\phi}_0(q,\omega'')\rangle$ gives us $\delta(p+q)$. Integrating over the $q$ and $\omega''$ variable we get
\begin{eqnarray}
\label{24}
\nonumber
\langle \tilde{\phi}_1(k,\omega)\tilde{\phi}_1(-k,-\omega)\rangle=G_0(k,\omega)\biggr[ \int \frac{\mathrm{d}p}{2\pi}\frac{\mathrm{d}\omega'}{2\pi} \frac{p^2}{|k-p|^{(2+\alpha)}}\\
\nonumber
G_0(p,\omega')G_0(-p,-\omega')\biggr] G_0(-k,-\omega)\\
\end{eqnarray}

The second term in Eq.(\ref{12}) has the same structure of $\langle \tilde{\phi}_1(k,\omega)\tilde{\phi}_1(-k,-\omega)\rangle$ while the third term has the same structure as
$\langle \tilde{\phi}_2(k,\omega)\tilde{\phi}_0(-k,-\omega)\rangle$.
Thus Eq.(\ref{22}) gives us 
\begin{equation}
\label{25}
\Sigma^2\sim\int \frac{\mathrm{d}p}{2\pi}\frac{\mathrm{d}\omega'}{2\pi} \frac{kp}{|k-p|^{(2+\alpha)}}G_0(p,\omega')
\end{equation}
or
\begin{equation}
\label{26}
\Sigma \sim k^{(1-\alpha)/2}
\end{equation}

while from Eq.(\ref{24}) we get
\begin{equation}
\label{27}
\Sigma \sim \int \frac{\mathrm{d}p}{2\pi}\frac{\mathrm{d}\omega'}{2\pi} \frac{p^2}{|k-p|^{(2+\alpha)}} \frac{1}{\omega'^2+D^2p^4}
\end{equation}

Since $\omega \sim \Sigma$, and neglecting $Dp^2$ term compared to $\Sigma$, power counting yields
\begin{equation}
\label{28}
\Sigma \sim \frac{k^{(1-\alpha)}}{\Sigma}
\end{equation}
which gives us the same result as in Eq.(\ref{26}).
We remark, in passing, that the result obtained in Eq.(\ref{26}) can also be obtained by introducing noise term in Eq.(\ref{6}). It should be noted that for a Kolmogorov like velocity field, $\alpha=-1/3$.

Thus, the effect of the non-linearity in $\mathcal{O}(\epsilon^2)$ is to replace the term $Dk^2$ by an effective diffusion term that looks like $\Sigma_0 k^{\beta}$.
We can, therefore, rewrite Eq.(\ref{6}) as
\begin{equation}
\label{29}
\frac{\partial \tilde{\phi}}{\partial t}=-\Sigma_0 k^{\beta} \tilde{\phi}
\end{equation}
with $\beta=(1-\alpha)/2$.
The two time correlation function $\langle \tilde{\phi}(k,\omega)\tilde{\phi}(-k,-\omega)\rangle$ becomes
\begin{equation}
\label{30}
\langle \tilde{\phi}(k,t_1)\tilde{\phi}(-k,t_2)\rangle=e^{-\Sigma_0 k^{\beta}(t_1+t_2)}
\end{equation}
The correlation $C(t_1,t_2)\equiv \langle \phi(x,t_1)\phi(x,t_2) \rangle$ for a fixed $x$ is given by
\begin{eqnarray}
\label{31}
\nonumber
C(t_1,t_2)&=&\int \mathrm{d}k \langle \tilde{\phi}(k,t_1)\tilde{\phi}(-k,t_2)\rangle\\
&=&\frac{1}{\beta}[\Sigma_0(t_1+t_2)]^{-1/\beta}
\end{eqnarray}
Define the normalized variable $X(t)=\phi(x,t)/\sqrt{\langle \phi^2(x,t)\rangle}$. Then, the correlation $\langle X(t_1)X(t_2)\rangle$ in terms of $C(t_1,t_2)$ becomes,
\begin{eqnarray}
\label{32}
\nonumber
\bar{C}(t_1,t_2) \equiv \langle X(t_1) X(t_2) \rangle &=&C(t_1,t_2)/\sqrt{C(t_1,t_1)C(t_2,t_2)}\\
\nonumber
&=& \biggr[ \frac{2\sqrt{t_1t_2}}{(t_1+t_2)}\biggr]^{1/\beta}\\
\end{eqnarray}
Making the usual transformation $\ln t=T$, Eq.(\ref{32}) becomes 
\begin{eqnarray}
\label{33}
\nonumber
\bar{C}(T_1,T_2)&=&\biggr[ \frac{2}{e^{1/2(T_1-T_2)}+e^{-1/2(T_1-T_2)}}\biggr]^{1/\beta}\\
\nonumber
&=&\bigr[\mathrm{Sech}(\frac{T_1-T_2}{2})\bigr]^{1/\beta}\equiv f(|T_1-T_2|)\\
\end{eqnarray}
The correlator in Eq.(\ref{33}) is now stationary since it depends only on the difference $|T_1-T_2|$ and non-Markovian. 

To determine the persistence exponent we adapt the method of IIA as explained in Ref \cite{2}. The basic assumption is that the intervals between the successive zeros of $X(T)$ are statistically independent. We will briefly outline the method here.
The first step is to construct the variable $\sigma=sign(X)$. The correlator $A(T)=\langle\sigma(T)\sigma(0) \rangle$ is given by
\begin{equation}
\label{34}
A(T)=\frac{2}{\pi}\arcsin[f(T)]
\end{equation}
If $p_n(T)$ be the probability that an interval of size $T$ contains $n$ zeros of $X(T)$, $P(T)$ be the distribution of intervals and $Q(T)$ be the probability that the left and right of the interval contains no zeros, then
\begin{eqnarray}
\label{35}
\nonumber
p_n(T)=\langle T \rangle^{-1}\int_0^T \mathrm{d}T_1\mathrm{d}T_2.....\mathrm{d}T_n Q(T_1)P(T_2-T_1)..\\
\nonumber
....P(T_n-T_{n-1})Q(T-T_n)\\
\end{eqnarray}
together with
\begin{equation}
\label{36}
A(T)=\sum_{n=0}^{\infty} (-1)^{n} p_n(T)
\end{equation}
where $\langle T \rangle=-2/A'(0)$.
Taking a Laplace transform of Eq.(\ref{35}) and using the fact that $\tilde{P}(s)=1-s\tilde{Q}(s)$, $\tilde{P}(s)$ and $\tilde{Q}(s)$ being the Laplace transform of $P(T)$ and $Q(T)$ respectively, we arrive at
\begin{eqnarray}
\label{37}
\nonumber
p_n(s)&=&\frac{1}{\langle T \rangle s^2}[1-\tilde{P}(s)]^2 \tilde{P}^{n-1}(s) \quad n\ge 1\\
&=& \frac{1}{\langle T \rangle s^2}[\langle T \rangle-1+\tilde{P}(s)].
\end{eqnarray}

Finally, substituting Eq.(\ref{37}) in Eq.(\ref{36}) and carrying out the summation over $n$ yields $\tilde{P}(s)$ in terms of $\tilde{A}(s)$, that is,
\begin{equation}
\label{38}
\tilde{P}(s)=[2-F(s)]/F(s).
\end{equation}
where $F(s)$ is given by
\begin{equation}
\label{39}
F(s)=1+\frac{\langle T \rangle s}{2}[1-s\tilde{A}(s)],
\end{equation}
$\tilde{A}(s)$ being the Laplace transform of $A(T)$.
For large $T$, $p_0(T)\sim e^{-\theta T}$ means that the exponent $\theta$ is given by the pole of $\tilde{P}(s)$ or the zero of $F(s)$.

In our present case, the correlator $A(T)$ is given by
\begin{equation}
\label{40}
A(T)=\frac{2}{\pi}\arcsin[\mathrm{Sech}(T/2)]^{1/\beta}.
\end{equation}
which gives $\langle T \rangle=\pi\sqrt{4\beta}$.
The function $F(s)$ has the form,
\begin{equation}
\label{41}
F(s)=1+\pi\sqrt{\beta}s[1-\frac{2}{\pi}s\int_0^{\infty}\mathrm{d}T e^{-sT} \arcsin([\mathrm{Sech}(T/2)]^{1/\beta})]
\end{equation}

The zeros of the function $F(s)$ are determined numerically. As a check for numerical verification we took the values $\alpha=-3,-1 \quad\textrm{and} -1/3$, which corresponds to $1/\beta=1/2,1$ and $3/2$  respectively. These values of $1/\beta$ correspond to the normal diffusion in $D=1,2$ and $3$ respectively. The exponents determined numerically using these three values of $1/\beta$ agrees well with the exponents reported in Ref[].
Finally, we have taken various values of $\alpha$ between $0$ and $1$ and have obtained the roots of $F(s)$ numerically. The obtained values of the exponents are listed below in Table A.
\\
\begin{center}
\bf{TABLE A}
\end{center}
\begin{center}
\begin{tabular}{|c|c|c|}
\hline
 $\alpha$ & $\beta=(1-\alpha)/2$& $\theta$ \\
\hline
0.1 &  2.22 & -0.29341041 \\
\hline
0.2 &  2.50 & -0.312802995 \\
\hline
0.3 &  2.86 & -0.336107784 \\
\hline
0.4 &  3.33 & -0.364881293 \\
\hline
0.5 &  4.00 & -0.401726555 \\
\hline
0.6 &  5.00 & -0.451442543 \\
\hline
0.7 &  6.67 & -0.524308324 \\
\hline
0.8 & 10.00 & -0.64860239 \\
\hline
 \end{tabular} 
\end{center}

In conclusion, we have considered the persistence phenomenon in advectecd passive scalar equation. In 1-dimension the velocity is drawn from a random distribution with $\langle v(k,\omega)v(-k,-\omega) \rangle \sim |k|^{-(2+\alpha)}$. The effect of the non-linearity is to replace the the zeroth order Green's function by it's general form $G^{-1}=-i\omega+Dk^2+\Sigma$, with $\Sigma \sim k^{\beta}$. We have determined the scaling form of $\Sigma$ using self-consistent theory, which gives $\beta =(1-\alpha)/2$. Thus, in $\mathcal{O}(\epsilon^2)$, the non-linearity replaces the original dynamics with an effective equation where the diffusion term due to molecular viscosity by a term of the form $\Sigma_0 k^{\beta}$. We have calculated the two time correlation for the effective process which has the form $[\mathcal{Sech}(T/2)]^{1/\beta}$. Finally, we have used IIA to calculate the persistence exponents.

{\bf{\large{Acknowledgement:}}}\\
D.C acknowledges Council for Scientific and Industrial Research, 
Govt. of India for financial support (Grant No.- 9/80(479)/2005-EMR-I).
D.C is also grateful to Prof. J.K.Bhattacharjee for many fruitfull discussions.

\end{document}